\documentclass[aps,preprint,preprintnumbers,amsmath,amssymb,nofootinbib]{revtex4}
\usepackage{amssymb}
\usepackage{lscape}
\usepackage{amsmath}
\usepackage{bm}
\usepackage{graphicx}
\usepackage{epsfig}
\begin{document}

\title{Large scale solitonic back-reaction effects in pre-inflation}
\author{$^{2}$ Juan Ignacio Musmarra\footnote{juani.musmarra@gmail.com}, $^{1,2}$ Mariano Anabitarte\footnote{anabitar@mdp.edu.ar},  $^{1,2}$ Mauricio Bellini
\footnote{{\bf Corresponding author}: mbellini@mdp.edu.ar} }
\address{$^1$ Departamento de F\'isica, Facultad de Ciencias Exactas y
Naturales, Universidad Nacional de Mar del Plata, Funes 3350, C.P.
7600, Mar del Plata, Argentina.\\
$^2$ Instituto de Investigaciones F\'{\i}sicas de Mar del Plata (IFIMAR), \\
Consejo Nacional de Investigaciones Cient\'ificas y T\'ecnicas
(CONICET), Mar del Plata, Argentina.}

\begin{abstract}
Using Relativistic Quantum Geometry (RQG), we study the emergence of back-reaction modes with solitonic properties, on astrophysical and cosmological scales, in a model of pre-inflation where the universe emerge from a topological phase transition. We found that, modes of the geometrical field that describes back-reaction effects related to larger scales (cosmological scales), are more coherent than those related to astrophysical scales, so that they can be considered a coarse-grained soliton.\\
\end{abstract}
\maketitle

\section{Introduction and motivation}

The existence of a pre-inflationary epoch with fast-roll of the inflaton field would introduce an
infrared depression in the primordial power spectrum. This depression might have left an imprint in the CMB anisotropy\cite{GR}.
It is supposed that during pre-inflation the universe begun to expand from some Planckian-size initial volume to thereafter pass to an inflationary epoch.
Some models consider the possibility of a pre-inflationary epoch in which the universe is dominated by radiation\cite{IC}. In this framework RQG should be very useful when we try to study the evolution of the geometrical
back-reaction effects given that we are dealing with Planckian energy scales, and back-reaction
effects should be very intense at these scales\cite{ab}. In a previous work\cite{pre} was suggested the metric
\begin{equation}\label{m}
d\hat{S}^2 =  \left(\frac{\pi a_0}{2}\right)^2 \frac{1}{\hat{\theta}^2} \left[{d\hat{\theta}}^2 + \hat{\eta}_{ij} d\hat{x}^i d\hat{x}^j\right],
\end{equation}
to describe the background space-time during pre-inflation. If we desire to describe an initially Euclidean 4D universe, that thereafter evolves to an asymptotic value $\hat{\theta} {\rightarrow } 0$, we must require $\hat{\theta}$ to be with an initial value $\hat\theta_0=\frac{\pi}{2}$. Furthermore, the nonzero components of the Einstein tensor, are
\begin{equation}
G_{00} = - \frac{3}{\hat{\theta}^2} , \qquad G_{ij} =  \frac{3}{\hat\theta^2 } \,\delta{ij},
\end{equation}
so that the radiation energy density and pressure, are respectively given in this representation by $\rho(\hat\theta) = \frac{1}{2\pi G} \frac{3}{(\pi a_0)^2}$, $P(\hat\theta) = - \frac{1}{4\pi G} \frac{3}{(\pi a_0)^2}$. The equation of state for the metric (\ref{m}), describes a vacuum expansion: $\omega(\hat\theta) =  - 1$. In this case the asymptotic scale factor, the Hubble parameter and the potential are respectively given by
\begin{equation}
a(t)= a_0\, e^{H_0 t}, \qquad \frac{\dot{a}}{a} = H_0 \qquad V= \frac{3}{8\pi G} H^2_0,
\end{equation}
so that the background field solution is given by a constant value: $\phi(t)= \phi_0$. This solution describes the field that drives a topological phase transition from a 4D Euclidean space to a 4D hyperbolic space-time.

In order to describe the exact back-reaction effects, we shall consider Relativistic Quantum Geometry (RQG), introduced in
\cite{rb}. In this formalism the manifold is defined with the connections\footnote{To simplify the notation we denote $\sigma_{\alpha} \equiv \sigma_{,\alpha}$.}
\begin{equation}\label{gama}
\Gamma^{\alpha}_{\beta\gamma} = \left\{ \begin{array}{cc}  \alpha \, \\ \beta \, \gamma  \end{array} \right\}+ \sigma^{\alpha} \hat{g}_{\beta\gamma} ,
\end{equation}
where $\delta{\Gamma^{\alpha}_{\beta\gamma}}=\sigma^{\alpha} \hat{g}_{\beta\gamma} $ describes the displacement of the Weylian manifold\cite{weyl} with respect to the Riemannian background, which is described by the Levi-Civita symbols in (\ref{gama}). In our approach, $\sigma(x^{\alpha})$ is a scalar field and the covariant derivative of the metric tensor in the Riemannian background manifold is null (we denote with a semicolon the Riemannian-covariant derivative): $\Delta g_{\alpha\beta}=g_{\alpha\beta;\gamma} \,dx^{\gamma}=0$. However, the Weylian covariant derivative\cite{weyl} on the manifold generated by (\ref{gama}) is nonzero: $ g_{\alpha\beta|\gamma} = \sigma_{\gamma}\,g_{\alpha\beta}$. From the action's point of view, the scalar field $\sigma(x^{\alpha})$ is a generic geometrical transformation that leaves invariant the action\cite{rb}
\begin{equation}\label{aac}
{\cal I} = \int d^4 \hat{x}\, \sqrt{-\hat{g}}\, \left[\frac{\hat{R}}{2\kappa} + \hat{{\cal L}}\right] = \int d^4 \hat{x}\, \left[\sqrt{-\hat{g}} e^{-2\sigma}\right]\,
\left\{\left[\frac{\hat{R}}{2\kappa} + \hat{{\cal L}}\right]\,e^{2\sigma}\right\},
\end{equation}
Hence, Weylian quantities will be varied over these quantities in a semi-Riemannian manifold so that the dynamics of the system preserves the action: $\delta {\cal I} =0$, and we obtain
\begin{equation}
-\frac{\delta V}{\hat{V}} = \frac{\delta \left[\frac{\hat{R}}{2\kappa} + \hat{{\cal L}}\right]}{\left[\frac{\hat{R}}{2\kappa} + \hat{{\cal L}}\right]}
= 2 \,\delta\sigma,
\end{equation}
where $\delta\sigma = \sigma_{\mu} dx^{\mu}$ is an exact differential and $\hat{V}=\sqrt{-\hat{ g}}$ is the volume of the Riemannian manifold. Of course, all the variations are in the Weylian geometrical representation, and assure us gauge invariance because $\delta {\cal I} =0$.
The metric that takes into account back-reaction effects on the background space-time (\ref{m}), is
\begin{equation}\label{mm}
d{S}^2 =  \left(\frac{\pi a_0}{2}\right)^2 \frac{1}{\hat{\theta}^2} \left[e^{2\sigma}\,{d\hat{\theta}}^2 + \hat{\eta}_{ij}\,e^{-2\sigma}\, d\hat{x}^i  d\hat{x}^j\right],
\end{equation}
with a determinant $V=\sqrt{-\hat{ g} }\,e^{-2\sigma}$. The back-reaction energy density fluctuations during inflation were calculated in a previous work, using RQG\cite{mb}
\begin{equation}
\frac{1}{\hat{\rho}} \frac{\delta \hat{\rho}}{\delta S} = - 2 \left(\frac{\pi}{2a_0}\right) \, \hat\theta \sigma',
\end{equation}
where the {\em prime} denotes the derivative with respect to $\hat\theta$. Here, ${\sigma}' = \left< (\sigma')^2 \right>^{1/2}$, such that
\begin{equation}
\left< (\sigma')^2 \right> = \frac{1}{(2\pi)^{3}} \, \int\, d^3k ({\xi}_k)' \, ({\xi}^*_k)'.
\end{equation}
The modes of the field $\sigma$: $\xi_k$, must be restricted by the normalization condition:
$({\xi}_k^*)' \xi_k - ({\xi}_k)' \xi^*_k= i \hat{\theta}^2 \left(\frac{2}{\pi a_0}\right)^2$, in order for the field $\sigma$ to be quantized\cite{rb}
\begin{equation}\label{con}
\left[\sigma(x), \sigma_{\mu}(y)\right] =i \, \hbar \Theta_{\mu} \delta^{(4)}(x-y).
\end{equation}
Here, $\Theta_{\mu}=\left[\hat{\theta}^2 \left(\frac{2}{\pi a_0}\right)^2,0,0,0\right]$ are the components of the background relativistic 4-vector on the Riemann manifold such that
\begin{equation}
\Theta_{\mu} \Theta^{\mu} =1.
\end{equation}
Notice that the commutator (\ref{con}) tends asymptotically to zero with the expansion of the universe, i.e., when $\hat{\theta} \rightarrow 0$. The equation that describes the dynamics of the
$\sigma$-modes, is
\begin{equation}\label{mm}
\xi_k'' -   \frac{2}{\hat\theta} \xi'_k + k^2\, \xi_k(\hat\theta) =0.
\end{equation}
The quantized solution of (\ref{mm}) results to be
\begin{equation}
\xi_k(\theta)= \frac{i}{2}\left(\frac{\pi}{2 a_0}\right) k^{-3/2} \,e^{-i k \hat\theta} \left[k\hat\theta-i\right].
\end{equation}
In this work we shall study the decoherence of the modes that describe the geometric back-reaction, on the cosmological and astrophysical sector of the spectrum. It is assumed that the modes that remains coherent during the primordial expansion of the universe, can be considered as a solitonic package that remains unaltered during pre-inflation, and later.
In these sectors the modes $\sigma_k(\hat\theta, \vec{r})$ are unstable; mainly on the cosmological sector of the spectrum. The range of wavelengths, of the astrophysical spectrum, is: $2\pi/(\epsilon k_0) > 2\pi/k > 2\pi/k_0$, such that the wave-number range, is
\begin{equation}\label{as}
\epsilon k_0 < k \leq k_0=\frac{\sqrt{2}}{\theta},
\end{equation}
such that $\epsilon \simeq 10^{-2.5}$. For smaller wave-numbers the spectrum describes cosmological scales.

\section{Energy density fluctuations from pre-inflation: Astrophysical versus cosmological scales}

It is supposed that during pre-inflation the universe begun to expand from some Planckian-size initial volume to thereafter pass to an inflationary epoch. In this framework RQG should be very useful when we try to study the evolution of the geometrical back-reaction effects given that we are dealing with Planckian energy scales, and back-reaction effects should be very intense at these scales. We are aimed to study solitonic back-reaction effects in the range of the spectrum which is today astrophysical, to be able to compare the coherence of the modes for that scale, with those of cosmological scales. In the astrophysical range of the spectrum the modes are slightly unstable. Furthermore, because we are describing an intermediate region of the spectrum, with an upper wave-number cut $k_{0}=\sqrt{2}/\hat\theta$. The range of validity is given in (\ref{as}), and the energy density fluctuations are given by
\begin{equation}
\left|\frac{1}{\hat{\rho}} \frac{\delta\hat{\rho}}{\delta{S}}\right| = \frac{\pi}{a_0} \hat{\theta} \left< \left(\sigma'\right)^2\right>^{1/2},
\end{equation}
where we denote with the {\em prime} the derivative with respect to $\hat{\theta}$. Furthermore, $\left< \left(\sigma'\right)^2\right>$ is
\begin{equation}
\left< \left(\sigma'\right)2\right> = \frac{1}{2\pi^2} \int^{k_{max}}_{k_0=\sqrt{2}/\hat{\theta}} dk\, k^2 \sigma'_k(\hat{\theta})\,\sigma'^*_k(\hat{\theta}).
\end{equation}
In other words, the expectation value of squared-$\sigma'$, calculated on the background metric provide us $(\delta\sigma')^2$, which means that we are considering the back-reaction effects as gaussian. The result expressed as a function of $\hat{\theta}$, is
\begin{equation}\label{ener}
\left|\frac{1}{\hat{\rho}} \frac{\delta\hat{\rho}}{\delta{S}}\right|_{Astro} = \frac{\pi}{4 \sqrt{2}\,a_0^2} \left(1-\epsilon^4\right)^{1/2}.
\end{equation}
These are the energy density fluctuations due to back-reaction effects (\ref{ener}), on the astrophysical range of the spectrum. We can compare this result with the energy density fluctuations on cosmological scales:
\begin{equation}\label{en1}
\frac{\left|\frac{1}{\hat{\rho}} \frac{\delta\hat{\rho}}{\delta{S}}\right|_{Astro}}{\left|\frac{1}{\hat{\rho}} \frac{\delta\hat{\rho}}{\delta{S}}\right|_{Cosmo}} =\frac{(1-\epsilon^4)^{1/2}}{\epsilon^2} \simeq  \epsilon^{-2}.
\end{equation}
This is a relevant result which assures that the ratio between astrophysical and cosmological energy density fluctuations is $10^5$.

\section{Large-scale quasi-coherent spectrum from pre-inflation}

Now we can rewrite the scalar field $\sigma(\vec{r},\hat{\theta})$ as a Fourier expansion in spherical $(r,\varphi,\phi)$-coordinates
\begin{equation}
\sigma(\vec{r},\hat{\theta}) = \int^{\infty}_{0} dk\,\sum_{lm}
\left[a_{klm} \,\bar\Phi_{klm}(\vec{r},\hat{\theta})+a^{\dagger}_{klm}\,
\bar\Phi^*_{klm}(\vec{r},\hat{\theta})\right],
\end{equation}
where $\bar\Phi_{klm}(\vec{r},\hat{\theta})= k^2 \,j_l\left(kr\right)
\bar\Phi_{kl}(\hat\theta) Y_{lm}(\varphi,\phi)$, $Y_{lm}(\varphi,\phi)$
being the spherical harmonics and $j_{l}(kr)$ the spherical Bessel
functions. If we consider that the annihilation and creation
operators obey the algebra
\begin{equation}
\left[a_{klm}, a^{\dagger}_{k'l'm'}\right] = \delta(k- k')
\delta_{ll'} \delta_{m m'}, \qquad \left[a_{klm},
a_{k'l'm'}\right]=\left[a^{\dagger}_{klm},
a^{\dagger}_{k'l'm'}\right]=0.
\end{equation}
Because the universe can be considered isotropic and homogeneous at large scales, we can study the spherically symmetric geometrical waves that emerge during pre-inflation. Hence, if we consider only the cases with $l=m=0$, we obtain that the scalar field $\sigma$ is
\begin{equation}
\sigma(\vec{r},\hat{\theta}) = \frac{1}{r}\,\int^{\infty}_{0} dk\,k^2\,\left[a_{k00} \, e^{i k(r+\|v_k(\hat\theta)\| \hat\theta)} +a^{\dagger}_{k00} \,e^{-i k(r+\|v_k(\hat\theta)\| \hat\theta)}\right],
\end{equation}
where we have considered that
\begin{equation}
\bar\Phi_{k00}(\vec{r},\hat{\theta}) =  k^2 \,\frac{e^{i k r}}{r} \,\xi_k(\hat\theta),
\end{equation}
and the phase velocity for some $k$-mode is
\begin{equation}
v_k(\hat\theta) = \frac{1}{k\hat\theta} \,{\rm ln}\left[ \xi_k(\hat\theta)\right],
\end{equation}
with squared norm $\|v_k(\hat\theta)\|^2= \Re\left[v_k(\hat\theta)\right]^2 + \Im\left[v_k(\hat\theta)\right]^2$. If we define $r=0$ when $\theta=\pi/2$, we can define the
shift phase: $\Omega_k(\hat\theta)$, of each $k$-mode, as
\begin{equation}
\Omega_k(\hat\theta) = k\left[\frac{\pi}{2} \left\|v_k(\pi/2)\right\| - \left\|v_k(\hat\theta)\right\| \hat\theta\right],
\end{equation}
such that when the phase transition begins, all the modes are in phase: $\Omega_k(\hat\theta=\pi/2)=0$. With this definition, modes with $\Omega_k(\hat\theta)=0$, for $\forall \,\hat\theta$, will be considered as coherent. In the Fig. (\ref{f1}) we have plotted modes with different wave-number $k$, which correspond to the astrophysical and cosmological sector of the spectrum. Notice that the modes with bigger $k$, are more shifted than those with smaller $k$, with the evolution of the universe. This means that those modes corresponding to the cosmological sector of the spectrum remain coherent during pre-inflation. These modes, with wavelengths $\lambda > 2\pi/(\epsilon \,k_0) $ can be considered with solitonic properties, and
the coarse-grained geometrical field $\sigma_{cg}(\hat\theta,\vec{r})$, can be expressed (in spherical coordinates), as
\begin{equation}
\sigma_{cg}\left(\hat\theta,\vec{r}\right) = \frac{1}{r} \,\int d^3k \, G\left(k,\hat\theta\right)\,\left[a_{k00} \, e^{i k(r+\|v_k(\hat\theta)\| \hat\theta)} +a^{\dagger}_{k00} \,e^{-i k(r+\|v_k(\hat\theta)\| \hat\theta)}\right],
\end{equation}
with the suppression function
\begin{equation}
G\left(k,\hat\theta\right) = \sqrt{\frac{1}{1+ \left(\frac{k_0(\hat\theta)}{k}\right)^N}},
\end{equation}
and $N$ to be experimentally determined.

\section{Final Comments}

\begin{figure}[h]
\noindent
\includegraphics[width=.6\textwidth]{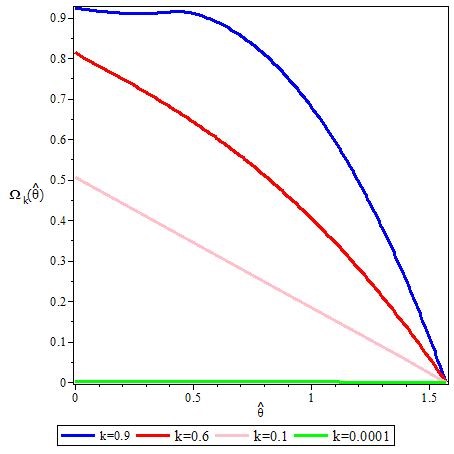}\vskip -0cm\caption{Plot of the shift phase $\Omega_k(\hat\theta)$, for different wavenumbers of the astrophysical and cosmological sector of the spectrum. The pre-inflationary epoch begins at $\hat\theta=\pi/2$ when all the modes are in phase. During pre-inflation, the modes corresponding to the cosmological sector remain coherent, but those of the astrophysical one, change their phase with respect to those of the cosmological sector. The mode with $k=k_0$ corresponds to $k=0.903$.}\label{f1}
\end{figure}

We have studied the properties of the modes related to back-reaction effects during pre-inflation in order to estimate the relative amplitude of energy density fluctuation in both, the astrophysical and cosmological ranges of the spectrum. These are $10^5$ times bigger, on astrophysical scales than on cosmological scales. However, on cosmological scales the modes of $\sigma$ remain coherent, meanwhile on astrophysical scales they shift their phase with respect to other modes of the spectrum. It is expected for this effect to be amplified on smaller scales. These results are shown in the Fig. (\ref{f1}). The mode with $k=k_0$ corresponds to $k=0.903$, with is plotted in blue. Notice that those modes with $k < k_0$ are unstable and remains more and more coherent as $k$ decrease, such that wavelengths in the range $\sqrt{2}\pi \hat\theta/\epsilon > \lambda > \sqrt{2}\pi \hat\theta$ can be considered as a classical soliton (with coherent modes that remain in phase). For this physical reason, the geometrical back-reaction field $\sigma$, can be considered as a coarse-grained (solitonic) scalar field on cosmological scales.

\newpage
\end{document}